\documentclass[prb,twocolumn,showpacs,amsmath,amssymb]{revtex4}
\usepackage[latin2]{inputenc}
\usepackage{amsmath}
\usepackage{graphicx}
\usepackage{dcolumn}
\usepackage{epstopdf}

\newcommand{\ba}{\begin{eqnarray*}}
\newcommand{\ea}{\end{eqnarray*}}
\newcommand{\baa}{\begin{eqnarray}}
\newcommand{\eaa}{\end{eqnarray}}
\def\bar{\begin{array}}
\def\ear{\end{array}}
\def\LB{\left(}
\def\RB{\right)}

\def\ra{\rightarrow}

\def\pr{^{\prime}}

\def\u{\uparrow}
\def\d{\downarrow}

\def\s{\sigma}
\def\f{\frac}

\def\ga{\gamma}

\begin{document}

\title{Kohn-Sham scheme for frequency dependent linear response}

\author{Ryan Requist}
\email{Ryan.Requist@physik.uni-erlangen.de}
\author{Oleg Pankratov}
\affiliation{
Theoretische Festk\"orperphysik, Friedrich-Alexander-Universit\"at Erlangen-N\"urnberg \\
Staudtstra\ss e 7-B2, 91058 Erlangen, Germany
}

\date{\today}

\begin{abstract}

We study the Kohn-Sham scheme for the calculation of the steady state linear response $\lambda n_{\omega}^{(1)}(\mathbf{r}) \cos \omega t$ to a harmonic perturbation $\lambda v^{(1)}(\mathbf{r}) \cos \omega t$ that is turned on adiabatically.  Although in general the exact exchange-correlation potential $v_{xc}(\mathbf{r},t)$ cannot be expressed as the functional derivative of a universal functional due to the so-called causality paradox, we show that for a harmonic perturbation the exchange-correlation part of the first-order Kohn-Sham potential $\lambda v_s^{(1)}(\mathbf{r}) \cos \omega t$ is given by $v_{xc}^{(1)}(\mathbf{r}) = \delta K_{xc}^{(2)}/\delta n_{\omega}^{(1)}(\mathbf{r})$. $K_{xc}^{(2)}$ is the exchange-correlation part of the second-order quasienergy $K_v^{(2)}$.  The Frenkel variation principle implies a stationary principle for $K_v^{(2)}[n_{\omega}^{(1)}]$.  We also find an analogous stationary principle and KS scheme in the time dependent extension of one-matrix functional theory, in which the basic variable is the one-matrix (one-body reduced density matrix).

\end{abstract}


\maketitle

\section{Introduction}

The description of a time dependent quantum state is fundamentally different from the description of a ground state because there is not a minimum principle for the former.  As density functional theory\cite{hohenberg1964} (DFT) is based on a minimum principle, it was not obvious that it could be extended to time dependent situations.  Yet such an extension, a time dependent DFT (TD DFT), was established by the Runge-Gross (RG) theorem,\cite{runge1984} which asserts that the time dependent density of a many-electron system determines the time dependent external potential up to an additive purely time dependent function, assuming a fixed initial state $\Psi(t_0) = \Psi_{t_0}$. The theorem applies to both interacting and noninteracting systems.  This is of great importance for applications because it implies that the density $n(\mathbf{r},t)$ of an interacting system can be reproduced by a noninteracting system with an effective potential $v_s(\mathbf{r},t)$, provided the initial state $\Phi(t_0) = \Phi_{t_0}$ of the noninteracting system is chosen to be compatible with $n(\mathbf{r},t)$.\cite{vanleeuwen1999} The potential $v_s(\mathbf{r},t)$, which is called the time dependent Kohn-Sham\cite{kohn1965} (KS) potential, is a functional of $n(\mathbf{r},t)$, $\Psi_{t_0}$ and $\Phi_{t_0}$.  Its exchange-correlation part $v_{xc}(\mathbf{r},t)$ can be defined from the equation
\begin{eqnarray}
v_s(\mathbf{r},t) = v(\mathbf{r},t) + v_H(\mathbf{r},t) + v_{xc}(\mathbf{r},t),
\end{eqnarray}
where $v_H(\mathbf{r},t)$ is the time dependent Hartree potential.  In contrast to $v_{xc}(\mathbf{r})$ in static DFT, the exact $v_{xc}(\mathbf{r},t)$ cannot be expressed as the functional derivative of a universal functional of $n(\mathbf{r},t)$.  This is a consequence of the so-called causality paradox.\cite{gross1996,vanleeuwen1998,mukamel2005,vignale2008}  In TD DFT, neither a general minimum principle nor even a stationary principle  has been found.\footnote{In Refs. \onlinecite{vanleeuwen1998} and \onlinecite{mukamel2005}, generalized action functionals were defined, but they were not required to be stationary.}  Such a principle might be helpful in the search for accurate approximations to $v_{xc}(\mathbf{r},t)$.  The quantum mechanical action principle does not lead to a stationary principle of the form $\delta A=0$, where $A$ is a functional of $n(\mathbf{r},t)$, because its density-functional formulation contains boundary terms.\cite{vignale2008}  

An important application of TD DFT is the calculation of dynamic polarizabilities and excitation energies.\cite{zangwill1980,ando1977,petersilka1996}  These can be obtained from the frequency dependent linear response function $\chi(\omega)=\chi(\mathbf{r},\mathbf{r}\pr;\omega)$.\footnote{In order to distinguish Fourier transform pairs such as $\chi(\mathbf{r},\mathbf{r}\pr;t-t\pr)$ and $\chi(\mathbf{r},\mathbf{r}\pr;\omega)$, we will use the following notations: $\chi(\omega)=\chi(\mathbf{r},\mathbf{r}\pr;\omega)$ and $\chi(t-t\pr)=\chi(\mathbf{r},\mathbf{r}\pr;t-t\pr)$.}  In the time domain, the retarded linear response function is defined as $\chi(\mathbf{r},t,\mathbf{r}\pr,t\pr) = \delta n(\mathbf{r},t)/\delta v(\mathbf{r}\pr,t\pr)$.  The KS system reproduces self-consistently the linear response of the interacting system to a perturbation $\delta v(\mathbf{r},t)$.  Thus, $\chi(\mathbf{r},t,\mathbf{r}\pr,t\pr)$ is related to the KS response function $\chi_s(\mathbf{r},t,\mathbf{r}\pr,t\pr) =\delta n(\mathbf{r},t)/\delta v_s(\mathbf{r}\pr,t\pr)$ by the Dyson-like equation\cite{gross1985}
\begin{align}
\chi(1,1\pr) &= \chi_s(1,1\pr) + \int d2 d3 \; \chi_s(1,2) \nonumber \\ &\times (v_c(\mathbf{r}_2,\mathbf{r}_3) \delta(t_2-t_3) +f_{xc}(2,3)) \chi(3,1\pr),\label{eqn:Dyson:time}
\end{align}
where $i=(\mathbf{r}_i,t_i)$, $v_c(\mathbf{r}_i,\mathbf{r}_j)=\left|\mathbf{r}_i-\mathbf{r}_j\right|^{-1}$ and $f_{xc}(i,j)=\delta v_{xc}(i)/\delta n(j)$ is the exchange-correlation kernel.  If the system is in its ground state at $t=t_0$, then $\chi(t,t\pr)$ depends only on the time difference $t-t\pr$ and not $t$ and $t\pr$ individually.  Thus, from the Fourier transform of (\ref{eqn:Dyson:time}) one obtains
\begin{align}
\chi(\mathbf{r},\mathbf{r}\pr;\omega) &= \chi_s(\mathbf{r},\mathbf{r}\pr;\omega) + \int d^3r_2 d^3r_3  \; \chi_s(\mathbf{r},\mathbf{r}_2;\omega) \nonumber \\ &\times (v_c(\mathbf{r}_2,\mathbf{r}_3) + f_{xc}(\mathbf{r}_2,\mathbf{r}_3;\omega)) \chi(\mathbf{r}_3,\mathbf{r}\pr;\omega) \label{eqn:Dyson:freq}
\end{align}
if $f_{xc}(\omega)$ exists.  Gross and Kohn state that ``For the most general situation, we do not know whether $f_{xc}$ exists..."\cite{gross1985}  Eq. (\ref{eqn:Dyson:freq}) implies the following formal representation:
\begin{eqnarray}
f_{xc}(\mathbf{r},\mathbf{r}\pr;\omega) = \chi_s^{-1}(\mathbf{r},\mathbf{r}\pr;\omega) - \chi^{-1}(\mathbf{r},\mathbf{r}\pr;\omega) - v_c(\mathbf{r},\mathbf{r}\pr). \label{eqn:fxc:freq}
\end{eqnarray}
Therefore, $f_{xc}(\omega)$ exists whenever $\chi(\omega)$ and $\chi_s(\omega)$ are invertible.  Although the RG theorem guarantees that $\chi(t,t\pr)$ is invertible (subject to the condition that the perturbation is analytic at $t=t_0$), the frequency dependent response function $\chi(\omega)$ is not always invertible.\cite{mearns1987}  As $\chi(\omega)$ is the Fourier transform of $\chi(t-t\pr)$, one might ask why the invertibility of $\chi(t-t\pr)$ does not imply the invertibility of $\chi(\omega)$.  We shall address this question in a later section.  The problem of the invertibility of the response functions is an instance of the $v$-representability problem in density functional theories.

In this paper, we introduce a stationary principle in TD DFT and use it to derive the KS equations for frequency dependent linear response.  Our approach is to work in the time domain, and by turning on a harmonic perturbation $\lambda v^{(1)}(\mathbf{r}) \cos \omega t$ adiabatically, we induce a steady state linear response density $\lambda n^{(1)}(\mathbf{r}) \cos \omega t$.  We prove that the quasienergy (an analog of the Bloch quasimomentum for systems periodic in time), to second order in $\lambda$, is a stationary functional of $n^{(1)}(\mathbf{r})$.  If the linear response density $\lambda n^{(1)}(\mathbf{r}) \cos \omega t$ can be reproduced by a KS system with an effective potential $v_s(\mathbf{r},t) = v_s^{(0)}(\mathbf{r}) + \lambda v_s^{(1)}(\mathbf{r}) \cos \omega t$, then the stationary principle implies that the exchange-correlation part of $v_s^{(1)}(\mathbf{r})$ is the functional derivative of the exchange-correlation part of the second-order quasienergy.  We also find an analogous stationary principle and linear response KS scheme in the time dependent extension of one-matrix functional theory\cite{gilbert1975} (1MFT).  In 1MFT, the external potential can be nonlocal in space and spin coordinates, i.e., it acts as an integral operator with the kernel $v(\mathbf{r} \sigma,\mathbf{r}\pr \sigma\pr)$.  The corresponding many-body operator is $\hat{V} = \sum_{\sigma \sigma\pr} \int d^3r d^3r\pr \hat{\psi}^{\dag}_{\sigma}(\mathbf{r}) v(\mathbf{r} \sigma,\mathbf{r}\pr \sigma\pr) \hat{\psi}_{\sigma\pr}(\mathbf{r}\pr)$.  Significantly, a theorem analogous to the RG theorem has not been found in TD 1MFT. 

TD DFT for the special case of time-periodic external potentials has been studied previously.\cite{bartolotti1982,deb1982}  However, the scope of these and later approaches,\cite{banerjee1997,aiga1999} which are based on a Hohenberg-Kohn-type minimum principle, are severely limited\cite{maitra2002,samal2006,maitra2007} because (i) the minimum principle is generally valid only for periodic potentials that have no Fourier component of frequency greater than the first excitation energy and (ii) one must assume the existence of Floquet states (reviewed below).  The approach we pursue here is distinct because it employs adiabatic ramping in real time and relies on a stationary principle instead of a minimum principle; hence, our results are valid for all frequencies (except, of course, resonance frequencies), and we need not assume the existence of Floquet states.  

The paper is organized as follows.  In Section \ref{sec:Floquet}, we review the basic properties of Floquet states, which are the fundamental states of a system with a time-periodic Hamiltonian.  The Frenkel variation principle is used to derive a stationary principle for the quasienergy of Floquet states obtained from adiabatic ramping.  In Section \ref{sec:PT}, we repeat the standard derivation of the frequency dependent linear response function, except that we employ an arbitrary adiabatic ramping function.  Theorems that establish a stationary principle and linear response KS scheme in TD DFT are proved in Section \ref{sec:TDDFT:KS}.  Analogous theorems in TD 1MFT are proved in Section \ref{sec:TD1MFT:KS}.  A simple illustration of the KS scheme in 1MFT is presented in Section \ref{sec:illustration}.  We comment briefly on the question of the existence of $f_{xc}(\omega)$ in Section \ref{sec:fxc}. 

\section{Stationary principle for the quasienergy\label{sec:Floquet}}

In this section, we derive a stationary principle for the quasienergy of Floquet states obtained by turning on adiabatically a time-periodic perturbation of the external potential.  

We begin by reviewing the properties of the wave function when the Hamiltonian is periodic in time, $\hat{H}(t+T)=\hat{H}(t)$.  If $\hat{H}(t)$ is an operator on a finite-dimensional Hilbert space, the Floquet theorem asserts that there exists a complete set of solutions of the form \cite{shirley1965,sambe1973}
\begin{equation}
\Psi_n(t) = \xi_n(t) e^{-i \epsilon_n t}, \qquad \xi_n(t+T)=\xi_n(t), \label{eqn:Floquet}
\end{equation}
which are called Floquet states or quasienergy eigenstates.  The quasienergy, $\epsilon_n$, is defined modulo $2\pi/T$.  The periodic factor $\xi_n(t)$ satisfies the equation
\begin{equation}
\big( \hat{H}(t) - i \partial_t \big) \xi_n(t) = \epsilon_n \xi_n(t). \label{eqn:Schrod:xi}
\end{equation}
If the system under consideration has an infinite-dimensional Hilbert space, the existence of Floquet states is not guaranteed by the Floquet theorem.  For many systems, it may be the case that $\hat{H}(t) - i \partial_t$ has no nontrivial eigenfunctions so that Floquet states do not exist.\cite{young1969}  

Now consider an $N$-electron system that starts in a ground state at $t=-\infty$ and experiences an adiabatically ramped (AR) periodic perturbation of the form $\lambda v^{(1)}_{\tau}(\mathbf{r},t) = \lambda f(t/\tau) v^{(1)}(\mathbf{r},t)$, where $v^{(1)}(\mathbf{r},t+T)=v^{(1)}(\mathbf{r},t)$ and $f=f(t/\tau)$ is a ramping function with time scale $\tau$. The many-body Hamiltonian is $\hat{H}_{\tau}(t) = \hat{H}^{(0)} + \lambda \hat{V}^{(1)}_{\tau}(t)$ with $\hat{H}^{(0)} = \hat{T} + \hat{W} + \hat{V}^{(0)}$, where $\hat{T}$ is the kinetic energy operator, $\hat{W}$ is the electron-electron interaction and $\hat{V}^{(0)} = \int d^3r \: v^{(0)}(\mathbf{r}) \hat{\psi}^{\dag}(\mathbf{r}) \hat{\psi}(\mathbf{r})$.  The ramping function is an arbitrary smooth function that satisfies $f(-\infty)=0$ and $f(\infty)=1$.  An example of a suitable ramping function is $(1+ \tanh(t/\tau))/2$.  The precise form of $f$ is inconsequential and will be left unspecified in the following.  The functions $v^{(1)}_{\tau}(\mathbf{r},t)$ form a one-parameter family of perturbations, and the action of an \textit{ideal} AR perturbation is realized by taking the limit $\tau \ra \infty$ at the end of the calculation.  Although $\hat{H}_{\tau}(t)$ is not exactly periodic, it is still possible to define a quasienergy if the system approaches a steady state in the limit $(t,\tau) \ra (\infty,\infty)$.   

Following Ref. \onlinecite{langhoff1972}, we factor the wave function as
\begin{equation}
\Psi(t) = \xi_{\tau}(t) \exp\big(-i \int_{-\infty}^t dt\pr K_{\tau}(t\pr)\big), \label{eqn:Psi:decomp}
\end{equation}
where 
\begin{equation}
K_{\tau}(t)=\f{\big< \xi_{\tau} \big| \hat{H}_{\tau}(t) - i \partial_t \big| \xi_{\tau} \big>}{\big< \xi_{\tau} \big|\xi_{\tau} \big>} . \label{eqn:K:t}
\end{equation} 
In accordance with the terminology of Ref. \onlinecite{langhoff1972},
the overall phase factor will be called the \textit{secular phase}, and the factor $\xi_{\tau}=\xi_{\tau}(t)$ will be called the \textit{nonsecular wave function}.  A system will be said to \textit{evolve adiabatically into a steady state} if the following two conditions are satisfied:  1) the nonsecular wave function tends to a unique function $\xi=\xi(t)$ with period $T$ in the limit $(t,\tau) \rightarrow (\infty,\infty)$, i.e., if for all $\epsilon>0$ there exist $t\pr$ and $\tau\pr$ such that $\left\| \xi_{\tau}(t)-\xi(t) \right\| < \epsilon$ when $t>t\pr$ and $\tau>\tau\pr$ and 2) all of the electrons remain localized in a finite region of space for all time.  Real systems, for which the spectrum usually has a continuum component, are not expected to evolve into such a steady state due to the possibility of multiquantum ionization.\cite{young1969}  However, if the perturbation is harmonic and the driving frequency $\omega = 2\pi/T$ is not a resonance frequency, the first-order term of the power series of $\xi_{\tau}$ with respect to $\lambda$ evolves adiabatically into a unique harmonic function $\xi^{(1)}$ (see Section \ref{sec:PT}).  In such cases, the system will be said to evolve adiabatically into a \textit{first-order steady state}.  The periodic function $\xi$, if it exists, will be called the \textit{steady state nonsecular wave function}, while $\xi^{(1)}$ will be called the \textit{first-order steady state nonsecular wave function}.  The quasienergy associated with $\xi$ is, cf. (\ref{eqn:Schrod:xi}),
\begin{equation}
K = \f{\big< \xi \big| \hat{H}(t) - i \partial_t \big| \xi \big>}{\big< \xi \big|\xi \big>}, \label{eqn:K}
\end{equation}
where $\hat{H}(t)$ is $\hat{H}_{\tau}(t)$ without the ramping function $f$.  

%We shall now use the Frenkel variation principle to show that the quasienergy is a stationary functional of $\xi$.\cite{langhoff1972}  
If the system evolves adiabatically into a steady state, the Frenkel variation principle implies that the quasienergy is a stationary functional of $\xi$.\cite{langhoff1972}  For an arbitrary time dependent Hamiltonian $\hat{H}\pr(t)$, the Frenkel variation principle states that $\Psi=\Psi(t)$ is the solution of the Schr\"odinger equation with the initial condition $\Psi(t_0)=\Psi_{t_0}$ if 
\begin{equation}
\big< \delta \Psi \big| \hat{H}\pr(t) - i \partial_t \big| \Psi \big> = 0, \label{eqn:Frenkel}
\end{equation}
where $\delta \Psi = \delta \Psi(t)$ is an arbitrary variation that satisfies $\delta \Psi(t_0)=0$.  Setting $\hat{H}\pr(t) = \hat{H}_{\tau}(t)$ and adding to (\ref{eqn:Frenkel}) its complex conjugate, we obtain
\begin{equation}
\delta \big< \Psi \big| \hat{H}_{\tau}(t) - i \partial_t \big| \Psi \big> + i \partial_t \big< \Psi \big| \delta \Psi \big> = 0.
\end{equation}
Substituting (\ref{eqn:Psi:decomp}) gives
\begin{equation}
\delta K_{\tau}(t) + i \f{\partial}{\partial t} \f{\big< \xi_{\tau} \big| \delta \xi_{\tau} \big>}{\big< \xi_{\tau} \big| \xi_{\tau} \big>} =0. \label{eqn:Frenkel2}
\end{equation}
As we are interested in the response to an AR periodic perturbation, we now consider a one-parameter family of variations $\delta \xi_{\tau}$ (with parameter $\tau$) such that $\xi_{\tau} + \delta \xi_{\tau}$ evolves adiabatically from the ground state to a steady state $\xi + \delta \xi$.  Then, upon taking the following limit and time average,\footnote{The double limit $\lim_{(t,\tau) \rightarrow (\infty,\infty)}$ must be taken with care.  It can be evaluated as $\lim_{\tau \ra \infty} \limsup_{t\ra \infty}$.} the second term of (\ref{eqn:Frenkel2}) vanishes, and we obtain
\begin{equation}
\lim_{(t,\tau) \rightarrow (\infty,\infty)} \f{1}{T} \int_t^{t+T} dt\pr
\; \delta K_{\tau}(t\pr) = 0, \label{eqn:Frenkel3}
\end{equation}
which suggests the definition 
\begin{eqnarray}
K_v[\xi] &=& \lim_{(t,\tau) \rightarrow (\infty,\infty)} \f{1}{T} \int_t^{t+T} dt\pr \f{\big< \xi_{\tau} \big| \hat{H}_{\tau}(t\pr) - i \partial_{t\pr} \big| \xi_{\tau} \big>}{\big< \xi_{\tau} \big| \xi_{\tau} \big>} \nonumber \\
&=& \f{1}{T} \int_t^{t+T} dt\pr \f{\big< \xi \big| \hat{H}(t\pr) - i \partial_{t\pr} \big| \xi \big>}{\big< \xi \big| \xi \big>},\label{eqn:def:Kv}
\end{eqnarray}
where the subscript $v$ denotes the given $v^{(1)}(\mathbf{r},t)$.  We define the domain of $K_v$ to be the space of all steady state nonsecular wave functions $\xi$ that can be obtained for some $v^{(1)}(\mathbf{r},t)$.  It is possible to choose a larger domain, but this is not necessary for our purposes.  Let $\xi_v$ denote the steady state nonsecular wave function corresponding to the given $v^{(1)}(\mathbf{r},t)$.  A variation $\delta \xi$ will be called an admissible variation if $\xi_v+\delta \xi$ is in the domain of $K_v$.  Thus, for every admissible variation $\delta \xi$, there exists a one-parameter family of variations $\delta \xi_{\tau}$ such that $\xi_{\tau} + \delta \xi_{\tau}$ evolves adiabatically into $\xi_v + \delta \xi$.  Hence, (\ref{eqn:Frenkel3}) implies the stationary principle 
\begin{eqnarray}
\delta K_v[\xi] = 0 \label{eqn:Jv:stat}
\end{eqnarray}
for an arbitrary admissible variation $\delta \xi$ at $\xi=\xi_v$.  In fact, this result follows from a direct calculation of the first variation, assuming only the existence of $\xi_v$.  We have carried out the above derivation based on the Frenkel variation principle because it will prove useful when Floquet states do not exist, a case to which we now turn. 

If the system does not evolve adiabatically into an exact steady state,  the limit in (\ref{eqn:def:Kv}) does not exist in general.  Nevertheless, for an AR nonresonant harmonic perturbation $\lambda v^{(1)}(\mathbf{r}) f(t/\tau) \cos \omega t$, the system evolves adiabatically into a first-order steady state described by $\xi_0^{(0)}+\lambda \xi^{(1)}$, where $\xi^{(0)}_0$ is the unperturbed ground state.  We now show that the second-order quasienergy is a stationary functional of $\xi^{(1)}$.  

Consider the trial function 
\begin{eqnarray}
\xi(r,t)  = \xi^{(0)}_0(r) + \lambda \xi^{(1)}(r,t), \label{eqn:trial}
\end{eqnarray}
where $r=(\mathbf{r}_1,\ldots, \mathbf{r}_N)$ and $\xi^{(1)}(r,t)$ is an arbitrary harmonic function with frequency $\omega$ subject to the constraint $\big<\xi^{(0)}_0\big|\xi^{(1)}\big>=0$.  Expanding (\ref{eqn:Frenkel2}) to second order in $\lambda$ and repeating the steps leading to (\ref{eqn:Jv:stat}), one obtains the stationary principle\cite{langhoff1972}
\begin{eqnarray}
\delta K^{(2)}_v[\xi^{(1)}] = 0, \label{eqn:stat:cond}
\end{eqnarray}
where
\begin{align}
K^{(2)}_v[\xi^{(1)}] &= \f{1}{T} \int_t^{t+T} dt\pr \Big( \big< \xi^{(1)} \big| \hat{H}^{(0)} - E_0^{(0)} - i \partial_{t\pr} \big| \xi^{(1)} \big> \nonumber \\ 
&+ \big< \xi^{(0)} \big| \hat{V}^{(1)}(t\pr) \big| \xi^{(1)} \big> + c.c. \Big). \label{eqn:Kv:2:simple}
\end{align}
Eq. (\ref{eqn:stat:cond}) applies for an arbitrary admissible variation $\delta\xi^{(1)}$ at $\xi^{(1)}=\xi^{(1)}_v$.  The trial wave function in (\ref{eqn:trial}), which is specified only up to first order in $\lambda$, is sufficient to obtain the quasienergy through third order,\footnote{This is a special case of the so-called $2n+1$ theorem.}
\begin{eqnarray}
K_v[\xi] = E^{(0)}_0 + \lambda^2 K^{(2)}_v[\xi^{(1)}] + \mathcal{O}(\lambda^4).
\end{eqnarray}
This is analogous to a well-known fact concerning the variational estimate of a ground state energy: the error in the energy is second order in the error of the trial wave function.  The first-order term of the trial function in (\ref{eqn:trial}) can be expressed as
\begin{eqnarray}
\xi^{(1)}(r,t) = \xi^{(1)}_+(r) e^{i \omega t} + \xi^{(1)}_-(r) e^{-i \omega t},
\end{eqnarray}
which leads to 
\begin{eqnarray}
K^{(2)}_v[\xi^{(1)}] &=& K^{(2)}_{v+}[\xi^{(1)}_+] + K^{(2)}_{v-}[\xi^{(1)}_-]; \nonumber \\
K^{(2)}_{v\pm}[\xi^{(1)}_{\pm}] 
&=& \big< \xi^{(1)}_{\pm} \big| \hat{H}^{(0)} - E^{(0)}_0 \pm \omega \big| \xi^{(1)}_{\pm} \big> \nonumber \\ 
&+& \big< \xi^{(0)}_0 \big| \hat{V}^{(1)} \big| \xi^{(1)}_{\pm} \big> + c.c., \label{eqn:Kv:2}
\end{eqnarray}
where $\hat{V}^{(1)}$ is the many-body operator corresponding to $v^{(1)}(\mathbf{r})$.  

In the above derivation, we made no assumptions about the higher order terms of the power series of $\xi_{\tau}$ with respect to $\lambda$.  
%The series to first order approaches a unique harmonic function in the limit $(t,\tau) \ra (\infty,\infty)$ if the driving frequency is nonresonant.  
The series truncated at order $N$,
\begin{eqnarray}
\xi_{\tau} \approx \xi_0^{(0)} + \lambda \xi_{\tau}^{(1)} + \cdots + \lambda^N \xi_{\tau}^{(N)},
\end{eqnarray}
may also approach a unique periodic function.  In most physical systems, there will exist an integer $N\pr$ such that the truncated series will cease to approach a periodic function if $N>N\pr$, owing to multiquantum resonances $N\omega = E_k-E_0$.  The analysis of this section can be extended up to the highest order for which there are no multiquantum resonances.

\section{Perturbation theory for adiabatically ramped harmonic perturbations \label{sec:PT}}

In order to show that the nonsecular wave function $\xi_{\tau}$, to first order in $\lambda$, evolves adiabatically into a unique harmonic function, we repeat the standard calculation of frequency dependent linear response.  We employ an arbitrary adiabatic ramping function $f=f(t/\tau)$ instead of the usual ramping function $e^{\eta t}$ with $\eta \rightarrow 0$.  Without loss of generality, we take the perturbation to be of the form $\lambda v^{(1)}_{\tau}(\mathbf{r},t)= \lambda  v^{(1)}(\mathbf{r}) f(t/\tau) \cos \omega t$ with $\omega \geq 0$.  The calculation will clarify the sense in which the steady state linear response is independent of the precise form of the ramping function in the adiabatic limit $\tau \rightarrow \infty$.

We consider an $N$-electron system with the Hamiltonian $\hat{H}_{\tau}(t)=\hat{H}^{(0)}+\lambda \hat{V}^{(1)}_{\tau}(t)$.  The system is assumed to start in the ground state at $t=-\infty$.  We expand the wave function as
\begin{equation}
\Psi(t) = \sum_k a_k(t) e^{-i E_k t} \Phi_k,
\end{equation}
where $\Phi_k$ and $E_k$ are the eigenstates and eigenenergies of $\hat{H}^{(0)}$.  The initial condition is $a_k(-\infty) = \delta_{k0}$. To first order (for $k\neq 0$),
\begin{eqnarray}
a_k^{(1)}(t) = -i V_{k0}^{(1)}\int_{-\infty}^t dt\pr f(t\pr/\tau) \cos(\omega t\pr) e^{i \omega_{k0} t\pr},
\end{eqnarray}
where $V_{k0}^{(1)} = \big<\Phi_k \big| \hat{V}^{(1)} \big| \Phi_0 \big>$ and $\omega_{k0} = E_k - E_0$.  Treating first the $e^{i\omega t}$ component of $\cos \omega t$, we obtain 
\begin{eqnarray}
a_{k+}^{(1)}(t) &=& -\f{i}{2} V_{k0}^{(1)} \int_{-\infty}^t dt\pr f(t\pr/\tau) e^{i (\omega + \omega_{k0}) t\pr} \nonumber \\
&=& -\f{i}{2} V_{k0}^{(1)} \left[ f(t\pr/\tau) \f{1}{i(\omega + \omega_{k0})} e^{i (\omega + \omega_{k0}) t\pr} \right]_{-\infty}^t \nonumber \\ &+& \f{i}{2} V_{k0}^{(1)} \int_{-\infty}^t dt\pr \f{df(t\pr/\tau)}{dt\pr} \f{1}{i(\omega + \omega_{k0})} e^{i (\omega + \omega_{k0}) t\pr} \nonumber
\end{eqnarray}
from integration by parts.  The last term vanishes in the limit $(t,\tau) \ra (\infty,\infty)$ if $\omega \neq -\omega_{k0}$, which can be shown as follows.  Let
\begin{eqnarray}
I(\tau) &=& \int_{-\infty}^{\infty} dt \f{df(t/\tau)}{dt} e^{i (\omega + \omega_{k0}) t} \nonumber \\
 &=& \int_{-\infty}^{\infty} ds \f{df(s)}{ds} e^{i (\omega + \omega_{k0}) \tau s},
\end{eqnarray}
where $s = t/\tau$.
According to the Riemann-Lebesgue lemma, $I(\tau)$ vanishes in the limit $\tau \ra \infty$ due to the rapidly oscillating phase factor.  The lemma requires only that the condition $\int_{-\infty}^{\infty} ds \left| df(s)/ds \right| < \infty$ is satisfied.  This determines the degree to which the ramping function is arbitrary:  the steady state linear response will be independent of the precise form of the ramping function provided only that the later satisfies $f(-\infty)=0$, $f(\infty)=1$, and the condition above.  This conclusion is  independent of the details of the system.  The same analysis applies for the $e^{-i \omega t}$ component of $\cos \omega t$ if $\omega \neq \omega_{k0}$.\footnote{More precisely, we assume there exists an $\epsilon>0$ such that none of the eigenvalues $E_k$ lie in the interval $\left[E_0+\omega-\epsilon,E_0+\omega+\epsilon\right]$.}  Combining the results, we find that the $a_k^{(1)}(t)$ approach the functions\footnote{The precise statement of the approach is $\lim_{\tau \ra \infty} \limsup_{t \ra \infty} \lim_{\lambda \ra 0} \left| (a_k(t) - \lambda \tilde{a}^{(1)}_k(t))/\lambda \right| = 0$, where the order of the limits is important.}
\begin{eqnarray}
\tilde{a}_k^{(1)}(t) &=& e^{i \omega_{k0}t} \f{V_{k0}^{(1)}}{2} \left[ \cos(\omega t) \left(\f{1}{\omega-\omega_{k0}} - \f{1}{\omega+\omega_{k0}}\right)\right. \nonumber \\ &-& \left. i \sin(\omega t) \left(\f{1}{\omega-\omega_{k0}} + \f{1}{\omega+\omega_{k0}}\right)\right].
\label{eqn:ak}
\end{eqnarray}
Thus, the first-order term of the nonsecular wave function approaches the harmonic function
\begin{eqnarray}
\xi^{(1)}(t) = \sum_{k \neq 0} \tilde{a}_k^{(1)}(t) e^{-i \omega_{k0} t} \Phi_k \label{eqn:xi:1}
\end{eqnarray}  
The corresponding linear response density is 
\begin{align}
n^{(1)}(\mathbf{r},t) &= \sum_{k\neq 0} 2\; \mathrm{Re}\left(\tilde{a}_k^{(1)}(t) e^{-i\omega_{k0}t}\right)  \big< \Phi_0 \big| \hat{n}(\mathbf{r}) \big| \Phi_k \big> \nonumber \\
&= \cos(\omega t) \sum_{k \neq 0} V_{k0}^{(1)} \left(\f{1}{\omega-\omega_{k0}} - \f{1}{\omega+\omega_{k0}}\right) \nonumber \\ 
&\times \big< \Phi_0 \big| \hat{n}(\mathbf{r}) \big| \Phi_k \big>, 
\end{align}
where $\hat{n}(\mathbf{r}) = \hat{\psi}^{\dag}(\mathbf{r}) \hat{\psi}(\mathbf{r})$ and we have used the fact that the $\Phi_k$ can be chosen real in the present case.  The frequency dependent linear response function is readily identified as
\begin{align}
\chi(\mathbf{r},\mathbf{r}\pr;\omega) &= \sum_{k \neq 0} \left(\f{1}{\omega-\omega_{k0}} - \f{1}{\omega+\omega_{k0}}\right)  \nonumber \\ &\times \big< \Phi_0 \big| \hat{n}(\mathbf{r}) \big| \Phi_k \big> \big< \Phi_k \big| \hat{n}(\mathbf{r}\pr) \big| \Phi_0 \big>.\label{eqn:chi}
\end{align}

\section{Kohn-Sham scheme in time dependent density functional theory\label{sec:TDDFT:KS}}

We now turn to the main results of the paper.  In this section, we prove two theorems that establish a stationary principle and KS scheme in TD DFT for the special case of harmonic perturbations.  

In the following, we assume that the frequency dependent density response function $\chi(\omega)$ is invertible.\footnote{As $\chi(\omega)$ is an operator between infinite dimensional function spaces, its invertibility can only be established with respect to these spaces.  We do not pursue this issue here, and we shall simply assume that $\chi$ is invertible on a sufficiently large space of linear response densities.}  This is not always true.  For example, $\chi(\omega)$ will not be invertible if it has any null eigenvalues apart from the trivial one corresponding to an arbitrary constant shift of the external potential.  Mearns and Kohn\cite{mearns1987} have given an explicit example in which $\chi(\omega)$ has nontrivial null eigenvalues; however, these occur only for particular isolated frequencies $\omega_p$.  The theorems below can be extended to such cases by requiring that $v^{(1)}_{\omega_p}(\mathbf{r})$ and $n^{(1)}_{\omega_p}(\mathbf{r})$ are orthogonal to all null ``directions.''  We also assume that $\omega \geq 0$ is not a resonance frequency, i.e., $\omega \neq (E_k-E_0)$. 

Consider an electron system that starts in a nondegenerate ground state at $t = -\infty$ and experiences an AR harmonic perturbation $\lambda v_{\omega}^{(1)}(\mathbf{r}) f(t/\tau) \cos \omega t$.  The system evolves adiabatically from the ground state with density $n^{(0)}(\mathbf{r})$ into a first-order steady state with density $n^{(0)}(\mathbf{r})+\lambda n_{\omega}^{(1)}(\mathbf{r}) \cos \omega t$.  Since, by assumption, $\chi(\omega)$ is invertible, the linear response density $n_{\omega}^{(1)}(\mathbf{r})$ determines the perturbation $v_{\omega}^{(1)}(\mathbf{r})$ up to an arbitrary constant. The perturbation, in turn, determines the first-order steady state nonsecular wave function $\xi^{(1)}$, as seen in (\ref{eqn:xi:1}).  Therefore, $\xi^{(1)}$ is a functional of $n_{\omega}^{(1)}(\mathbf{r})$ on the space of linear response densities that can be obtained for some $v_{\omega}^{(1)}(\mathbf{r})$.  This space will be referred to as the linear response $v$-representable (LR VR) space.  Let $n^{(1)}_{\omega,v}(\mathbf{r})$ be the linear response density corresponding to the given $v^{(1)}_{\omega}(\mathbf{r})$.  We may now state the first theorem.

\textbf{\textit{Theorem 1.}} --- The second-order quasienergy $K^{(2)}_v$ is a functional of $n^{(1)}_{\omega}(\mathbf{r})$, and it satisfies the stationary condition $\delta K^{(2)}_v = 0$ for an arbitrary admissible variation $\delta n^{(1)}_{\omega}(\mathbf{r})$ at $n^{(1)}_{\omega}(\mathbf{r})=n^{(1)}_{\omega,v}(\mathbf{r})$.  

\textbf{\textit{Proof.}} --- The second-order quasienergy is a functional of $n^{(1)}_{\omega}=n^{(1)}_{\omega}(\mathbf{r})$ by composition of $K^{(2)}_v[\xi^{(1)}]$ and $\xi^{(1)}[n^{(1)}_{\omega}]$.  Therefore, the domain of $K^{(2)}_v[n^{(1)}_{\omega}]$ is the LR VR space.  A variation $\delta n^{(1)}_{\omega}(\mathbf{r})$ is admissible if $n_{\omega,v}^{(1)}(\mathbf{r}) + \delta n_{\omega}^{(1)}(\mathbf{r})$ is LR VR.  In Section \ref{sec:Floquet}, it was shown that $K^{(2)}_v[\xi^{(1)}]$ is stationary for an arbitrary admissible variation $\delta \xi^{(1)}$ at $\xi^{(1)} = \xi^{(1)}_v$.  As $\xi^{(1)}$ is a functional of $n_{\omega}^{(1)}$ on the LR VR space, there exists an admissible $\delta \xi^{(1)}$ corresponding to every admissible $\delta n^{(1)}_{\omega}$.  Therefore, $K^{(2)}_v[n^{(1)}_{\omega}]$ is stationary because $\delta K^{(2)}_v = (\delta K^{(2)}_v/\delta \xi^{(1)}) (\delta \xi^{(1)}/\delta n^{(1)}_{\omega}) \delta n^{(1)}_{\omega}$, suppressing coordinate integrations, and the first factor is zero.  

This stationary principle can be used to derive an expression for the exchange-correlation potential of a linear response KS system.  The linear response KS system is a noninteracting system that experiences the potential $v_s(\mathbf{r},t)=v_s^{(0)}(\mathbf{r})+\lambda v_{s}^{(1)}(\mathbf{r}) f(t/\tau) \cos \omega t$ and reproduces the first-order steady state density $n^{(0)}(\mathbf{r})+\lambda n_{\omega}^{(1)}(\mathbf{r}) \cos \omega t$ of the interacting system.  Such a system will exist if the following two $v$-representability conditions are satisfied.  
%at $t=-\infty$    % that experiences an AR harmonic perturbation   %+\mathcal{O}(\lambda^2)

\textbf{Condition (1a).} The ground state density of the interacting system is noninteracting v-representable (VR-N).  This means that there exists a system of noninteracting electrons with a potential $v_s^{(0)}(\mathbf{r})$ such that the ground state is nondegenerate and reproduces the ground state density of the interacting system.  

\textbf{Condition (1b).} The frequency dependent response function $\chi_s(\omega)$ of the noninteracting system in condition (1a) is invertible on the LR VR space of the interacting system. 

\textbf{\textit{Theorem 2.}} --- If an interacting system satisfies conditions (1a) and (1b), then the exchange-correlation part of $v_{s}^{(1)}(\mathbf{r})$ is given by
\begin{equation}
v^{(1)}_{xc}(\mathbf{r}) = \f{\delta K_{xc}^{(2)}}{\delta n^{(1)}_{\omega}(\mathbf{r})}, \label{eqn:vxc}
\end{equation}
where $K_{xc}^{(2)}[n^{(1)}_{\omega}]$ is the exchange-correlation part of the second-order quasienergy. 

\textbf{\textit{Proof.}} --- In analogy with Ref.~\onlinecite{runge1984}, the exchange-correlation part of the second-order quasienergy is defined as
\begin{align}
K_{xc}^{(2)}[n^{(1)}_{\omega}] &= \f{1}{T} \int_t^{t+T} dt\pr \big< \xi^{(1)}  \big| \hat{W} \big| \xi^{(1)}  \big>  + S_{W}^{(2)}[n^{(1)}_{\omega}] \nonumber \\
&- S_{0}^{(2)}[n^{(1)}_{\omega}] - \f{1}{2} \int \! d^3r d^3r\pr  \f{n^{(1)}_{\omega}(\mathbf{r}) n^{(1)}_{\omega}(\mathbf{r}\pr)}{\left| \mathbf{r} - \mathbf{r}\pr \right|}, \label{eqn:Kxc}
\end{align}
where
\begin{eqnarray}
S_{W}^{(2)}[n^{(1)}_{\omega}] = \f{1}{T} \int_t^{t+T} \!dt\pr \big< \xi^{(1)} \big| \hat{T} + \hat{V}^{(0)} - E_0^{(0)} - i \partial_{t\pr} \big| \xi^{(1)} \big> \nonumber
\end{eqnarray}
and
\begin{eqnarray}
S_{0}^{(2)}[n^{(1)}_{\omega}] = \f{1}{T} \int_t^{t+T} \!dt\pr \big< \xi^{(1)}_s \big| \hat{T} + \hat{V}^{(0)} - E_0^{(0)} - i \partial_{t\pr} \big| \xi^{(1)}_s \big>, \nonumber 
\end{eqnarray}
and $\xi^{(1)}_s$ is the first-order steady state nonsecular wave function of the KS system.  The last term of (\ref{eqn:Kxc}) subtracts the Hartree contribution, $K_H^{(2)}$.  The above functionals are universal in the sense that they do not depend on $v^{(1)}_{\omega}(\mathbf{r})$.  However, they do depend on the ground state density and the value of $\omega$, though this dependence will not be indicated explicitly.  With these definitions, one finds $K_v^{(2)} = S_0^{(2)} + \int d^3r\: v_{\omega}^{(1)}(\mathbf{r}) n_{\omega}^{(1)}(\mathbf{r}) + K_H^{(2)} + K_{xc}^{(2)}$.  Hence, the stationary condition of theorem 1 gives
\begin{align}
\f{\delta K_v^{(2)}}{\delta n_{\omega}^{(1)}(\mathbf{r})} &= \f{\delta S_0^{(2)}}{\delta n_{\omega}^{(1)}(\mathbf{r})} + v^{(1)}_{\omega}(\mathbf{r}) + v^{(1)}_H(\mathbf{r}) + v^{(1)}_{xc}(\mathbf{r})\nonumber \\ 
&= 0, \label{eqn:Euler}
\end{align}
where $v^{(1)}_H(\mathbf{r})=\delta K_H^{(2)}/\delta n^{(1)}_{\omega}(\mathbf{r})$.   The KS system also evolves into a first-order steady state, and the stationary condition for its second-order quasienergy 
$K_v^{(2)} = S_0^{(2)} + \int d^3r\: v_{s}^{(1)}(\mathbf{r}) n_{\omega}^{(1)}(\mathbf{r})$ will be identical to (\ref{eqn:Euler}) if 
\begin{eqnarray}
v_{s}^{(1)}(\mathbf{r}) = v_{\omega}^{(1)}(\mathbf{r}) + v_H^{(1)}(\mathbf{r}) + v_{xc}^{(1)}(\mathbf{r}).
\end{eqnarray}
Since $\chi_s(\omega)$ is invertible, $v_{s}^{(1)}(\mathbf{r})$ is uniquely defined. Thus, $v_{xc}^{(1)}(\mathbf{r})$ is given by (\ref{eqn:vxc}).  This completes the proof.

The steady state linear response density of the interacting system can be computed from the expression $n^{(1)}(\mathbf{r},t) = \sum_{i=1}^{N} \phi_i^{*(0)}(\mathbf{r}) \phi_i^{(1)}(\mathbf{r},t) + c.c.$, where $\phi_i^{(0)}(\mathbf{r})$ are the ground state KS orbitals and $\phi_i^{(1)}(\mathbf{r},t)$ are obtained from first-order perturbation theory for the single-particle Schr\"odinger equation
\begin{eqnarray}
i \partial_t \phi_i(\mathbf{r},t) &=& \big( -\f{1}{2} \nabla^2 + v_s^{(0)}(\mathbf{r}) \big) \phi_i(\mathbf{r},t)\nonumber \\ &+& \lambda v_{s}^{(1)}(\mathbf{r}) f(t/\tau) \cos(\omega t) \phi_i(\mathbf{r},t)
\end{eqnarray}
with the initial condition $\phi_i(\mathbf{r},-\infty) = \phi_i^{(0)}(\mathbf{r})$.  We remark that theorem 2 can be extended to cases where the ground state density is not VR-N but rather EVR-N (noninteracting ensemble $v$-representable) by following an approach analogous to the one taken in the next section.

Theorem 2 implies that the exchange-correlation kernel can be calculated as
\begin{equation}
f_{xc}(\mathbf{r},\mathbf{r}\pr;\omega) = \f{\delta^2 K_{xc}^{(2)}}{\delta n_{\omega}^{(1)}(\mathbf{r}) \delta n_{\omega}^{(1)}(\mathbf{r}\pr)}.
\end{equation}
The stationary principle for the second-order quasienergy does not entail a causality paradox because its basic variable $n_{\omega}^{(1)}(\mathbf{r})$ is time independent.  It is straightforward to derive the following formal expression:
\begin{align}
K_v^{(2)} &= - \f{1}{2} \int dr dr\pr n_{\omega}^{(1)}(\mathbf{r}) \chi^{-1}(\mathbf{r},\mathbf{r}\pr;\omega) n_{\omega}^{(1)}(\mathbf{r}\pr) \nonumber \\ 
&+ \int dr n_{\omega}^{(1)}(\mathbf{r}) v_{\omega}^{(1)}(\mathbf{r}).
\end{align}

%The second-order quasienergy functional bears a close resemblance to the action functional $A[n]= \int_{t_1}^{t_2} dt \left< \Psi(t) \left| i \partial_t - H(t) \right| \Psi(t) \right>$ introduced by Runge and Gross.\cite{runge1984}  It avoids the causality paradox because its basic variable $n_{\omega}^{(1)}(\mathbf{r})$ is time independent. 

\section{Kohn-Sham scheme in time dependent one-matrix functional theory\label{sec:TD1MFT:KS}}

In this section, we generalize the theorems of the previous section to TD 1MFT. 

Before proceeding, it will be helpful to review some basic results from static 1MFT.  The defining feature of 1MFT is that it has the capacity to treat systems in which the external potential is nonlocal with respect to the space and spin coordinates.  Accordingly, the necessary basic variable is the one-matrix (one-body reduced density matrix), which is defined as 
\begin{equation}
\gamma(x,x\pr) = N \int dx_2 \ldots dx_N  \rho(x, x_2, \ldots x_N ; x\pr,x_2, \ldots x_N), 
\label{eqn:1-matrix}
\end{equation}
where $x=(\mathbf{r},\s)$, $\int dx = \sum_{\s} \int d^3r$, and 
$\hat{\rho} = \sum_n w_n \left| \Psi_n \right> \left< \Psi_n \right|$
is the $N$-electron density matrix with ensemble weights $w_n$ such 
that $\sum_n w_n = 1$.  By extending the Hohenberg-Kohn theorem,\cite{hohenberg1964} Gilbert proved\cite{gilbert1975} that 1) the one-matrix uniquely determines the ground state wave function and 2) there is a universal energy functional $E_v[\ga]$ that attains its minimum at the ground state one-matrix.  There is also a KS scheme in 1MFT.  From the stationary condition for the energy functional, Gilbert derived the equation 
\begin{eqnarray}
-\f{1}{2} \nabla^2 \phi_i(x) + \int dx\pr v_s(x,x\pr) \phi_i(x\pr) = \epsilon_i \phi_i(x), \label{eqn:1MFT-KS}
\end{eqnarray}
where $v_s(x,x\pr) = v(x,x\pr)+ \delta W/\delta \gamma(x\pr,x)$ and $W=W[\gamma]$ is the universal electron-electron interaction functional.  This equation can be interpreted as the single-particle Schr\"odinger equation for the orbitals of a noninteracting system (the 1MFT KS system).  The potential $v_s(x,x\pr)$ is a functional of the one-matrix.  The ground state one-matrix of the interacting system can be obtained by solving self-consistently (\ref{eqn:1MFT-KS}) together with
\begin{equation}
\gamma(x,x\pr) = \sum_i f_i \phi_i(x) \phi_i^*(x\pr),
\label{eqn:1-matrix:diag}
\end{equation}
where $f_i$ are occupation numbers that satisfy $\sum_i f_i = N$ and $0 \leq f_i \leq 1$.   Generally, fractional occupation numbers are required to reproduce the one-matrix of the interacting system, not only the values $0$ and $1$ as in DFT.  This scheme was originally described as paradoxical\cite{gilbert1975,valone1980,tal1985} because the stationary condition implies that essentially all of the $\epsilon_i$ collapse to a single level.  Therefore, it appeared that the single-particle Schr\"odinger equation would not define unique orbitals. However, when the occupation numbers are shifted slightly from their ground state values, the KS equations have a self-consistent solution for a one-matrix that is close to the ground state one-matrix and for which the degeneracy is lifted.\cite{requist2008}  Thus, the correct ground state orbitals can be obtained in the limit that the occupation numbers approach their ground state values.  The ground state orbitals, which are called \textit{natural orbitals}, are the eigenfunctions of the ground state one-matrix, and the corresponding eigenvalues are the occupation numbers.\cite{loewdin1955}  As the occupation numbers are fractional, it is useful to interpret the KS system as adopting an ensemble state. 
 
In the time dependent version of 1MFT, a general KS scheme has not been found.  Such a scheme should have the capacity to treat systems in which the time dependent external potential is nonlocal with respect to the space and spin coordinates.  In TD DFT, the existence of the KS scheme is implied by the RG theorem.  But in TD 1MFT, it is not known whether there is a theorem as general as the RG theorem, i.e., for as general a class of time dependence.  The Bogoliubov-Born-Green-Kirkwood-Yvon hierarchy provides an equation of motion for the one-matrix.  However, this equation contains the two-matrix (two-body reduced density matrix), and it is not known whether the two-matrix is a universal functional of the one-matrix when the external potential is nonlocal.  As we are interested in frequency dependent linear response, we shall narrow our attention to the special case of AR harmonic perturbations.  In this case, there is a KS scheme in TD 1MFT.  Theorems 3 and 4 below are generalizations of theorems 1 and 2.

We shall need to refer to a different linear response function.  The \textit{one-matrix response function} in the time domain is defined as $\chi(x_1,x_1\pr,t_1;x_2,x_2\pr,t_2) = \delta \gamma(x_1,x_1\pr,t_1)/\delta v(x_2,x_2\pr,t_2)$.  The frequency dependent one-matrix response function will be denoted $\chi(\omega) =\chi(x_1,x_1\pr;x_2,x_2\pr;\omega)$.  Similarly, the one-matrix response function of the KS system will be denoted $\chi_{s}(\omega)$. In order to emphasize the analogy between this section and the previous section, some of the notations will be duplicated.
  
As the KS system can be interpreted as adopting an ensemble state,  our first step will be to prove a stationary principle for an ensemble-weighted quasienergy.

Consider an electron system that experiences an AR local or nonlocal perturbation $\lambda v_{\omega}^{(1)}(x,x\pr) f(t/\tau) \cos \omega t$.  Suppose that the system starts at $t=-\infty$ in the ensemble state $\hat{\rho}^{(0)} = \sum_n w_n \left| \Phi_n \right> \left< \Phi_n \right|$, where $\Phi_n$ are orthogonal stationary states, no two of which are in resonance, i.e., $E_m-E_n\neq \omega$ for all $m$ and $n$.  Let $\gamma_{\omega,v}^{(1)}(x,x\pr)$ denote the steady state linear response one-matrix corresponding to the given $v_{\omega}^{(1)}(x,x\pr)$.  Also, let $K_n^{(2)}$ denote the second-order quasienergy that would be obtained if the system were to start in the pure stationary state $\Phi_n$ instead of the mixed state $\hat{\rho}^{(0)}$.  It is convenient to introduce a notation in which Hermitian functions of $(x,x\pr)$ are expressed with respect to the complete basis of ground state natural orbitals $\phi_i^{(0)}(x)$.  For example, $v_{\omega}^{(1)}(ij) = \int dx dx\pr \phi_i^{*(0)}(x) v_{\omega}^{(1)}(x,x\pr) \phi_j^{(0)}(x\pr)$.  We may now state the stationary principle.

\textbf{\textit{Theorem 3.}} --- If the frequency dependent response function of the ensemble is invertible for a given $\omega$, then the ensemble-weighted second-order quasienergy 
\begin{eqnarray}
K^{(2)}_v = \sum_n w_n K_n^{(2)} \label{eqn:Kv:weighted}
\end{eqnarray}
is a functional of $\gamma^{(1)}_{\omega}=\gamma^{(1)}_{\omega}(ij)$ and satisfies the stationary condition $\delta K^{(2)}_v = 0$ for an arbitrary admissible variation $\delta \gamma^{(1)}_{\omega}$ at $\gamma^{(1)}_{\omega}=\gamma^{(1)}_{\omega,v}$.    

\textbf{\textit{Proof.}} --- Let $\xi^{(1)}_n$ be the first-order steady state nonsecular wave function that would be obtained if the system were to start in the stationary state $\Phi_n$ at $t=-\infty$.  The ensemble-weighted second-order quasienergy, for fixed $w_n$, is a functional of $\gamma^{(1)}_{\omega}$ because each $K^{(2)}_n$ is a functional of $\xi^{(1)}_n$ and each $\xi^{(1)}_n$ is a functional of $\gamma^{(1)}_{\omega}$.  Each $K^{(2)}_n$ can be shown to be a functional of $\xi^{(1)}_n$ by repeating the arguments leading to (\ref{eqn:Kv:2}) for a system initially in the state $\Phi_n$, assuming, as we have, that $\Phi_n$ is not in resonance with any of the other stationary states.  Each $\xi^{(1)}_n$ is a functional of $\gamma^{(1)}_{\omega}$ because $\gamma^{(1)}_{\omega}$ determines $v^{(1)}_{\omega}$ (up to a constant), which, in turn, determines $\xi^{(1)}_n$.  The first variation of $K^{(2)}_v[\gamma^{(1)}_{\omega}]$ can be expressed as $\delta K^{(2)}_v = \sum_n w_n (\delta K^{(2)}_n/\delta \xi^{(1)}_n) (\delta \xi^{(1)}_n/\delta \gamma^{(1)}_{\omega}) \delta \gamma^{(1)}_{\omega}$, where the coordinate integrations have been suppressed.  A variation $\delta \gamma^{(1)}_{\omega}$ is admissible if $\gamma^{(1)}_{\omega,v}+\delta \gamma^{(1)}_{\omega}$ is in the LR VR space of the ensemble.  The first variation vanishes for an arbitrary admissible variation $\delta \gamma^{(1)}_{\omega}$ because $\delta K^{(2)}_n/\delta \xi^{(1)}_n = 0$ for all $n$, which follows from a straightforward extension of the arguments in Sec.~\ref{sec:Floquet}. 

Now consider an electron system that starts in a nondegenerate ground state at $t=-\infty$ and experiences an AR local or nonlocal perturbation $\lambda v_{\omega}^{(1)}(x,x\pr) f(t/\tau) \cos \omega t$.  If the following two conditions are satisfied, then there is a linear response KS system in TD 1MFT.

\textbf{Condition (2a).} The ground state one-matrix is noninteracting ensemble $v$-representable (EVR-N).  This means that there exists a system of noninteracting electrons with a potential $v_s^{(0)}(ij)$ such that the ground state, which may be an ensemble state $\hat{\rho}_s^{(0)}$, reproduces the one-matrix of the interacting system. 

\textbf{Condition (2b).} The frequency dependent one-matrix response function $\chi_s(\omega)$ of the noninteracting system in condition (2a) is invertible on the space of all $\gamma^{(1)}_{\omega}$ that (i) are LR VR in the interacting system and (ii) have no diagonal and degenerate components. 

The \textit{diagonal components} of are simply the linear response occupation numbers $f_i^{(1)}=\gamma^{(1)}_{\omega}(ii)$, while the \textit{degenerate components} are $\gamma^{(1)}_{\omega}(jk)+\gamma^{(1)}_{\omega}(kj)$ and $-i\gamma^{(1)}_{\omega}(jk)+i\gamma^{(1)}_{\omega}(kj)$, where $\phi^{(0)}_j$ and $\phi^{(0)}_k$ are any pair of occupationally degenerate natural orbitals, i.e., natural orbitals for which $f_j^{(0)}=f_k^{(0)}$.  The diagonal and degenerate components correspond to null eigenfunctions of $\chi_s(\omega)$, so they are not LR VR in the KS system.  Therefore, the appropriate basic variable for the linear response KS system is $\overline{\gamma}^{(1)}_{\omega}=\overline{\gamma}^{(1)}_{\omega}(x,x\pr)=\sum_{ij}\pr \overline{\gamma}^{(1)}_{\omega}(ij) \phi_i^{(0)}(x) \phi_j^{*(0)}(x\pr)$, where the prime indicates that the diagonal and degenerate components are excluded from the sum.  In effect, $\overline{\gamma}^{(1)}_{\omega}$ describes the orbital degrees of freedom but not the occupation numbers.  Similarly, let $\overline{v}_{\omega}^{(1)}$ denote the projection of the given perturbation $v_{\omega}^{(1)}$ to the nondiagonal and nondegenerate subspace.  Also, let $\overline{\gamma}_{\omega,v}^{(1)}$ be the linear response corresponding to $\overline{v}_{\omega}^{(1)}$.

\textbf{\textit{Theorem 4.}} ---  If an interacting system satisfies conditions (2a) and (2b), then its first-order steady state one-matrix $\gamma^{(0)}(ij)+\lambda \overline{\gamma}_{\omega}^{(1)}(ij) \cos \omega t$ can be reproduced by a KS system with the potential $v_s(ij,t)=v_s^{(0)}(ij)+\lambda \overline{v}_{s}^{(1)}(ij) f(t/\tau) \sin \omega t$.  The contribution to $\overline{v}_{s}^{(1)}(ij)$ from the electron-electron interaction is given by
\begin{equation}
\overline{w}^{(1)}(ij) = \f{\delta K^{(2)}_{int}}{\delta \overline{\gamma}_{\omega}^{(1)}(ji)},
\end{equation}
where $K^{(2)}_{int}[\overline{\gamma}_{\omega}^{(1)}]$ is the interaction part of the second-order quasienergy. 

\textbf{\textit{Proof.}} --- The proof is analogous to the proof of theorem 2.  The existence of $\overline{v}_s(ij,t)$ follows from conditions (2a) and (2b).  According to condition (2a), the ground state one-matrix can be reproduced by a KS system in the ensemble state $\rho_s^{(0)}=\sum_n w_{s,n} \left|\Phi_{s,n} \right>\left<\Phi_{s,n}\right|$.  The $\Phi_{s,n}$ can be taken to be Slater determinants of $N$ natural orbitals.  To show that $\overline{w}^{(1)}(ij)$ is the functional derivative of a universal interaction functional, we first define
\begin{eqnarray*}
S_{W}^{(2)}[\overline{\gamma}^{(1)}_{\omega}] \!\!\! &= \f{1}{T} \int_t^{t+T} \!\!\! & \!dt\pr \big< \xi^{(1)} \big| \hat{T} + \hat{V}^{(0)} - E_0^{(0)} - i \partial_{t\pr} \big| \xi^{(1)} \big>, \\
S_{0}^{(2)}[\overline{\gamma}^{(1)}_{\omega}] \!\!\! &= \f{1}{T} \int_t^{t+T} \!\!\! & \!dt\pr \sum_n w_{s,n} \nonumber \\ 
&&\times \big< \xi^{(1)}_{s,n} \big| \hat{T} + \hat{V}^{(0)} - E_0^{(0)} - i \partial_{t\pr} \big| \xi^{(1)}_{s,n} \big>, \\
&= \f{1}{T} \int_t^{t+T} \!\!\! & \!dt\pr \sum_i f_i^{(0)} \nonumber \\ 
&&\times \big< \phi_i^{(1)} \big| \hat{t} + \hat{v}^{(0)} - E_0^{(0)} - i \partial_{t\pr} \big| \phi_i^{(1)} \big>,
\end{eqnarray*}
where $\hat{t}$ and $\hat{v}^{(0)}$ are one-body operators and $\xi^{(1)}_{s,n}$ is the first-order steady state nonsecular wave function that would be obtained if the KS system were to start in the pure state $\Phi_{s,n}$.  The contribution to the second-order quasienergy from the electron-electron interaction is
\begin{eqnarray}
K_{int}^{(2)}[\overline{\gamma}^{(1)}_{\omega}] &=& \f{1}{T} \int_t^{t+T} dt\pr \big< \xi^{(1)}  \big| \hat{W} \big| \xi^{(1)}  \big>  + S_{W}^{(2)}[\overline{\gamma}^{(1)}_{\omega}] \nonumber \\
&-& S_{0}^{(2)}[\overline{\gamma}^{(1)}_{\omega}]. 
\end{eqnarray}
We have not partitioned $K_{int}^{(2)}$ into Hartree and exchange-correlation terms because the linear response density $n_{\omega}^{(1)}$, which appears in the Hartree term, cannot be expressed in terms of $\overline{\gamma}_{\omega}^{(1)}$ alone, for it depends also on diagonal and degenerate components of $\gamma_{\omega}^{(1)}$.  In terms of the above functionals, the second-order quasienergy can be written $K_v^{(2)} = S_0^{(2)} + \sum_{ij} \overline{v}_{\omega}^{(1)}(ij) \overline{\gamma}_{\omega}^{(1)}(ji) + K_{int}^{(2)}$.  Hence, the stationary condition (theorem 3) for the interacting system is
\begin{eqnarray}
\delta K_v^{(2)} &=& \sum_{ij} \Big[ \f{\delta S_0^{(2)}}{\delta \overline{\gamma}_{\omega}^{(1)}(ji)}  + \overline{v}_{\omega}^{(1)}(ij) + \overline{w}^{(1)}(ij) \Big] \delta \overline{\gamma}_{\omega}^{(1)}(ji) \nonumber \\ 
&=& 0 \label{eqn:Euler:gamma}
\end{eqnarray}
for an arbitrary admissible variation $\delta \overline{\gamma}_{\omega}^{(1)}(ij)$.  A variation $\delta \overline{\gamma}_{\omega}^{(1)}(ij)$ is admissible if $\overline{\gamma}_{\omega,v}^{(1)}(ij)+\delta \overline{\gamma}_{\omega}^{(1)}(ij)$ can be obtained for some $\overline{v}_{\omega}^{(1)}(ij)$.  In order for $\delta K_v^{(2)}$ to vanish for an arbitrary admissible variation, the expression in brackets in (\ref{eqn:Euler:gamma}) must vanish for all ``directions" except the diagonal and degenerate directions.  Eq. (\ref{eqn:Euler:gamma}) is identical to the stationary condition for the ensemble-weighted second-order quasienergy of a KS system\footnote{The condition that the states of the ensemble are mutually nonresonant is automatically satisfied if $f_i^{(0)} \neq 0,1$ for all $i$, which is the typical case, due to the total degeneracy of the KS system.} with the potential $v_s(ij,t)=v_s^{(0)}(ij)+\lambda \overline{v}_{s}^{(1)}(ij) f(t/\tau) \sin \omega t$, if 
\begin{eqnarray}
\overline{v}_{s}^{(1)}(ij) = \overline{v}_{\omega}^{(1)}(ij) + \overline{w}^{(1)}(ij).
\end{eqnarray}

We remark that the KS perturbation in TD 1MFT must be advanced by a phase of $\pi/2$ with respect to the given perturbation because the linear response $\lambda \overline{\gamma}_{\omega}^{(1)}(ij) \cos \omega t$ of the KS system has a phase delay.  As the KS system can be interpreted as adopting an ensemble state, its one-matrix is governed by the equation of motion
\begin{eqnarray}
i \partial_t \hat{\gamma}(t) = \left[\hat{t} + \hat{v}_s(t),\hat{\gamma}(t)\right]. \label{eqn:motion}
\end{eqnarray}
which gives, to first order in $\lambda$, 
\begin{eqnarray}
i \big<\phi_i^{(0)}|\dot{\phi}_j^{(1)}(t)\big> = \big<\phi_i^{(0)}\big| \hat{\overline{v}}_{s}^{(1)}(t)\big|\phi_j^{(0)}\big>, \label{eqn:orbital}
\end{eqnarray}
for any pair of natural orbitals for which $f_i^{(0)} \neq f_j^{(0)}$.  Thus, the steady state linear response of the interacting system can be computed from the expression
\begin{eqnarray}
\overline{\gamma}_{\omega}^{(1)}(ij,t) &=& f_j^{(0)} \big<\phi_i^{(0)} \big| \phi_j^{(1)}(t)\big> + f_i^{(0)} \big<\phi_i^{(1)}(t) \big| \phi_j^{(0)}\big> \nonumber \\
&=& \f{i}{\omega} (f_j^{(0)}-f_i^{(0)}) \overline{v}_s^{(1)}(ij) \cos \omega t.
\end{eqnarray}  
While the linear response KS scheme does not give the diagonal and degenerate components of $\gamma^{(1)}_{\omega}(ij)$, they can be obtained instead by finding the stationary point of $K_v^{(2)}[\gamma_{\omega}^{(1)}]$.\footnote{If any occupation number $f_i^{(0)}$ is equal to $0$ or $1$, then $\chi(\omega)$ will have a corresponding null vector.\cite{requist2008}  Although a condition of theorem 3, namely that $\chi(\omega)$ is invertible, would therefore be violated, the stationary principle remains valid for variations in the complement of the null space of $\chi(\omega)$.}  This is analogous to the situation in static 1MFT, where the occupation numbers, which are not determined directly by the KS equations, can be obtained from the minimization of the energy.\cite{requist2008}  

The linear response KS scheme implies the Dyson-like equation
\begin{eqnarray}
\overline{\chi}(ij;kl;\omega) &=& \overline{\chi}_s(ij,kl;\omega) + \sum_{mnpq}  \;\overline{\chi}_s(ij,mn;\omega) \nonumber \\ &\times& \overline{\Lambda}(mn,pq;\omega)  \overline{\chi}(pq,kl;\omega), \label{eqn:1MFT:Dyson}
\end{eqnarray}
where $\overline{\Lambda}(\omega) = \delta \overline{w}^{(1)}/\delta \overline{\gamma}_{\omega}^{(1)}$.  
Excitation energies can be calculated from the poles of the response function by the method proposed in Ref. \onlinecite{petersilka1996}.  However, if the potential is local, it may be preferable to use (\ref{eqn:Dyson:freq}) rather than (\ref{eqn:1MFT:Dyson}) because the single-particle eigenvalues of the DFT KS system are often good approximations to the exact low-lying spectrum, while the 1MFT KS system provides no approximation at all due to its total degeneracy.  If the potential is nonlocal, the inverse response function, which also contains information about the excitations, can be obtained from the second functional derivative of $K_v^{(2)}$, as seen from the following expression:
\begin{align}
K_v^{(2)} &= -\f{1}{2} \int dx_1 dx_1\pr dx_2 dx_2\pr \: \gamma_{\omega}^{(1)}(x_1\pr,x_1) \nonumber \\ 
&\chi^{-1}(x_1,x_1\pr;x_2,x_2\pr;\omega) \gamma_{\omega}^{(1)}(x_2,x_2\pr) \nonumber \\
&+ \int dx_1 dx_1\pr \: v_{\omega}^{(1)}(x_1,x_1\pr) \gamma_{\omega}^{(1)}(x_1\pr,x_1).
\end{align}

\section{Illustrative example\label{sec:illustration}}

In the previous section, it was found that the time dependent KS scheme in 1MFT does not directly determine the linear response of the occupation numbers.  To clarify this aspect of the theory, we use the KS scheme to calculate the linear response in a simple example. 

We consider a simple version of the Hubbard model.\cite{fradkin1991}  The electrons are confined to a discrete lattice, each site of which can accommodate up to two electrons.  The electron-electron interaction is modeled by an on-site interaction.  As a further simplification, we consider that there are only two sites and only two electrons.\footnote{The static version of this model was treated by 1MFT in Ref.~\onlinecite{requist2008}.}  The unperturbed Hamiltonian is
\begin{eqnarray}
\hat{H}^{(0)} &=& - \tilde{t} \sum_{\s} \big( c_{1\s}^{\dag} c_{2\s} + c_{2\s}^{\dag} c_{1\s} \big) \nonumber \\
&+& U \LB \hat{n}_{1\u} \hat{n}_{1\d} + \hat{n}_{2\u} \hat{n}_{2\d} \RB +\hat{V}^{(0)},
\end{eqnarray}
where $c_{i\s}^{\dag}$ and $c_{i\s}$ are the creation and annihilation operators of an electron at site $i$ with spin $\s$ and $\hat{n}_i=\sum_{\s} c_{i\s}^{\dag} c_{i\s}$.  The first term of the Hamiltonian represents the kinetic energy by introducing ``hopping'' between the sites with energy parameter $\tilde{t}$.

For $\hat{V}^{(0)}=0$, the ground state is $\left|\Phi_0\right> = (1/\sqrt{2}) \big( y c_{1\u}^{\dag} c_{1\d}^{\dag} + x c_{1\u}^{\dag} c_{2\d}^{\dag} + x c_{2\u}^{\dag} c_{1\d}^{\dag} + y c_{2\u}^{\dag} c_{2\d}^{\dag} \big) \left| 0\right>$, where $x=\cos(\pi/4-\alpha_0/2)$ and $y=\sin(\pi/4-\alpha_0/2)$ with $\tan \alpha_0 = U/4\tilde{t}$.  

We are interested in the linear response to a nonlocal perturbation. For simplicity we shall consider only spin independent perturbations, so there will be only spatial nonlocality.  Thus, the AR perturbation is $\lambda \hat{V}^{(1)} f(t/\tau) \cos \omega t$ with
\begin{eqnarray}
\hat{V}^{(1)}  &=& \sum_{ij\sigma} v_{ij}^{(1)} c_{i\sigma}^{\dag} c_{j\sigma} \nonumber \\
&=& \sum_{\alpha} v_{\alpha}^{(1)} \hat{\sigma}_{\alpha},
\end{eqnarray}
where we have introduced Pauli operators, e.g., $\hat{\sigma}_x = \sum_{\sigma} (c_{1\sigma}^{\dag} c_{2\sigma} + c_{2\sigma}^{\dag} c_{1\sigma})$.  The spatial one-matrix of a general state $\Psi$ is defined as
\begin{equation}
 \gamma(ij) = \sum_{\sigma} \big< \Psi \big| c_{j\sigma}^{\dag} c_{i\sigma} \big| \Psi \big>,
 \label{eqn:1-matrix:lattice}
\end{equation}
which can be expressed in terms of the natural orbitals as
\begin{eqnarray}
\gamma(ij) = \sum_{k} f_{k} \phi_{k}(i) \phi_{k}^*(j). \label{eqn:1-matrix:discrete}
\end{eqnarray}
There are only two natural orbitals,
\begin{eqnarray}
\phi_a &=& \LB \begin{array}{rr} \cos(\theta/2) e^{-i \psi/2} \\ \sin(\theta/2) e^{i \psi/2}\end{array} \RB \quad \mathrm{and} \nonumber \\
\phi_b &=& \LB \begin{array}{rr} \sin(\theta/2) e^{-i \psi/2} \\ -\cos(\theta/2) e^{i \psi/2} \end{array} \RB. 
\label{eqn:orbitals}
\end{eqnarray}
As a Hermitian $2\times 2$ matrix, the spatial one-matrix can be expressed as
\begin{eqnarray}
\gamma &=&  I + A \LB \sin \theta \cos \psi \sigma_x + \sin \theta \sin \psi \sigma_y + \cos \theta \sigma_z \RB \nonumber \\
 &=&  I + \vec{\gamma} \cdot \vec{\sigma}; \quad \vec{\gamma} = (\gamma_x,\gamma_y,\gamma_z);
 \label{eqn:1-matrix:param}
\end{eqnarray}
where $A=(f_a-f_b)/2$ and $\vec{\sigma}$ is the vector of Pauli matrices.  It is also convenient to express the one-matrix response function $\chi(\omega)$ with respect to the Pauli basis, e.g., $\chi_{xy} = \delta \gamma_x/\delta v_y$.  For the ground state $\Phi_0$, we obtain
\begin{eqnarray}
\chi(\omega) = 8 \!\!\LB \begin{array}{ccc} \f{\omega_{30}}{\omega^2-\omega_{30}^2} (x^2-y^2)^2 & 0 & 0 \\ 0 & \f{\omega_{20}}{\omega^2-\omega_{20}^2} x^2 & i \f{\omega}{\omega^2-\omega_{20}^2} xy \\ 0 & -i \f{\omega}{\omega^2-\omega_{20}^2} xy & \f{\omega_{20}}{\omega^2-\omega_{20}^2} y^2 \end{array} \RB, \nonumber
\end{eqnarray}
where $\omega_{k0}=E_k-E_0$.  The Kohn-Sham response function 
\begin{eqnarray}
\chi_s(\omega) = \f{4A}{\omega} \LB \begin{array}{ccc} 0 & 0 & 0 \\ 0  & 0 & i \\ 0 & -i & 0 \end{array} \RB \nonumber
\end{eqnarray}
has one null vector corresponding to the null linear response of the occupation numbers to a ``diagonal" perturbation $\delta v \left|\phi_a \right>\left< \phi_a \right| - \delta v\left|\phi_b \right>\left< \phi_b \right|$.  

For $\hat{V}^{(0)}\neq 0$, the KS response function becomes
\begin{eqnarray}
\chi_s(\omega) = \f{4A}{\omega}  \!\!\LB \begin{array}{ccc} 0 & i \cos \theta & -i \sin \theta \sin \psi \\ -i \cos \theta & 0 & i \sin \theta \cos \psi \\ i \sin \theta \sin \psi & -i \sin \theta \cos \psi & 0 \end{array} \RB \nonumber.
\end{eqnarray}
This has the null vector $(\sin \theta \cos \psi, \sin \theta \sin \psi, \cos \theta)$, which is just the unit vector with polar angle $\theta$ and azimuthal angle $\psi$.  We observe that this vector is parallel to the $\vec{\gamma}$ of the ground state, cf. (\ref{eqn:1-matrix:param}), which implies that any perturbation of the KS system, even a nonlocal perturbation, can change only the direction of $\vec{\gamma}$ and not its magnitude.  The magnitude $|\vec{\gamma}|=A$ is related to the difference of the occupation numbers (the sum is fixed, $f_a+f_b=2$).  As noted in Section \ref{sec:TD1MFT:KS}, 
the occupation numbers of the KS orbitals are not changed, to first order, by \textit{any} perturbation to the KS system.  This is a general feature of the KS scheme in 1MFT.  Nevertheless, the linear response of the occupation numbers can be obtained from the stationary condition $\delta K_v^{(2)}=0$.

\section{On the existence of $f_{xc}(\omega)$\label{sec:fxc}}

In this section, we explain why the invertibility of $\chi(t,t\pr)$, which is established by the RG theorem, does not imply the invertibility of $\chi(\omega)$ for a pure frequency component.  

The RG theorem implies that the inverse response function $\chi^{-1}(t,t\pr)$ is defined on the space $\mathcal{N}(\Psi_{t_0},t_0)$, which consists of all $n^{(1)}(\mathbf{r},t)$ that can be realized for the given initial state $\Psi(t_0) = \Psi_{t_0}$ by some perturbation that is analytic at $t=t_0$.  In order to obtain $\chi^{-1}(\omega)$ from the Fourier transform of $\chi^{-1}(t,t\pr)$, we must have $\chi^{-1}(t,t\pr)=\chi^{-1}(t-t\pr)$.  This will be the case only if the system is in the ground state (or a stationary state) of the unperturbed Hamiltonian at $t=t_0$.  Therefore, the relevant space is $\mathcal{N}(\Psi_{gs},t_0)$, where $\Psi_{gs}$ is the ground state.  The time $t_0$ is arbitrary but \textit{finite}.  Thus, it can be shown that the invertibility of $\chi(t-t\pr)$ implies the invertibility of $\chi(\omega)$ on the space $\mathcal{N}_{\omega}(\Psi_{gs},t_0)$, which consists of all ${n^{(1)}(\mathbf{r},\omega)}$ that are the Fourier transform of some $n^{(1)}(\mathbf{r},t) \in \mathcal{N}(\Psi_{gs},t_0)$.  However, $\mathcal{N}_{\omega}(\Psi_{gs},t_0)$ is too small to establish the invertibility of $\chi(\omega)$ for a pure frequency component $\omega$.  In other words, it does not contain the elements $n^{(1)}(\mathbf{r}) (\delta(\omega-\Omega) + \delta(\omega+\Omega))/2$, corresponding to $n^{(1)}(\mathbf{r},t) = n^{(1)}(\mathbf{r}) \cos \Omega t$.  Such elements are absent because a system in a perfect steady state with density $n_{gs}(\mathbf{r}) + \lambda n^{(1)}(\mathbf{r}) \cos \Omega t + \mathcal{O}(\lambda^2)$ for all time is generally never in an \textit{instantaneous} ground state.  Therefore, there is no time $t_0$ at which to specify the initial condition as required above.  Hence, the invertibility of $\chi(\omega)$ and the existence of $f_{xc}(\omega)$ are not implied by the RG theorem.

\section{Conclusions}

One of the fundamental questions that can be asked about a quantum system is:  How does it respond to a harmonic perturbation?  The first-order response of the density to a weak perturbation is described by the linear response function $\chi(\mathbf{r},\mathbf{r}\pr;\omega)$.  Considerable effort has been devoted to calculating $\chi(\mathbf{r},\mathbf{r}\pr;\omega)$ with the Dyson-like equation (\ref{eqn:Dyson:freq}).   This equation contains the exchange-correlation kernel $f_{xc}(\omega)$.  In this paper, we have shown that $f_{xc}(\mathbf{r},\mathbf{r}\pr;\omega) = \delta^2 K_{xc}^{(2)}/(\delta n_{\omega}^{(1)}(\mathbf{r}) \delta n_{\omega}^{(1)}(\mathbf{r}\pr))$, where $K_{xc}^{(2)}[n_{\omega}^{(1)}]$ is a universal functional.  

The RG theorem establishes the existence of time dependent KS equations, but it has not been possible to derive the exchange-correlation potential from a stationary principle.  The quantum mechanical action principle does not provide a suitable stationary principle because its density-functional formulation contains boundary terms.\cite{vignale2008}  For the special case of harmonic perturbations, we have found a stationary principle for the quasienergy that can be used to derive the first-order exchange-correlation potential.   

If the external potential of a time dependent system is nonlocal, then it is not known whether a KS scheme exists in general.  Although 1MFT has the scope to treat nonlocal potentials, a theorem as general as the RG theorem has not been found in TD 1MFT.  By extending the stationary principle for the quasienergy to TD 1MFT, we have shown that there is a KS scheme for the linear response of the natural orbitals to a harmonic perturbation.  The KS system experiences an adiabatically ramped perturbation of the form $v_s(x,x\pr,t) = v_s^{(0)}(x,x\pr) + \lambda v_s^{(1)}(x,x\pr) \sin \omega t$.  The part of $v_s^{(1)}(x,x\pr)$ due to interactions can be calculated from the functional derivative of a universal functional.  In contrast to the DFT KS system, the linear response of the 1MFT KS system has a phase delay of $\pi/2$, so that the KS potential must be advanced by a phase of $\pi/2$ with respect to the given external potential.

\bibliography{bibliography}
\end{document}